# Fuzzy propositional logic associated with quantum computational gates

Graciela Domenech *[1] and Hector Freytes [2]

April 28, 2005

1. Instituto de Astronomía y Física del Espacio (IAFE)
Casilla de Correo 67, Sucursal 28, 1428 Buenos Aires, Argentina
2. Escuela de Filosofía - Universidad Nacional de Rosario,
Entre Ríos 758, 2000, Rosario, Argentina

**Abstract**

We apply residuated structures associated with fuzzy logic to develop certain aspects of information processing in quantum computing from a logical perspective. For this purpose we introduce an axiomatic system whose natural interpretation is the irreversible quantum Poincaré structure.

## 1 INTRODUCTION

This paper is framed in a wide program of the application of fuzzy logic to analyze quantum information. In [10], we presented a logical formalization associated with contextuality using Heyting structures of the intuitionistic calculus. They are a particular case of residuated lattices [14], which are the algebraic counterpart of some systems of fuzzy logic. In the same line, we apply here residuated structures to develop certain aspects of information processing in quantum computing from a logical perspective.

Recently, increasing attention has been paid to quantum computational logic (QCL) [6] [8] [12], a logic related with quantum computation [18] [15],

*Fellow of the Consejo Nacional de Investigaciones Científicas y Técnicas (CONICET)

different from QL. On the one hand, QCL differs from quantum logic (QL) first introduced by Birkhoff and von Neumann [4] in that it is not an orthomodular structure. On the other hand, they have different semantics. A sentence in QL may be interpreted as a closed subspace of the Hilbert space of the system, the space of its possible pure states. Instead, the meaning of an elementary sentence in QCL is a quantum information quantity encoded in a collection of qbits (unit vectors pertaining to the tensorial product of two dimensional complex Hilbert spaces, that represent pure states) or in qmixes (positive semi-definite Hermitian operators of trace one on that space, that represent statistical mixtures of pure states). Moreover, conjunction and disjunction have to be interpreted otherwise than in QL because they are not associated to join and meet lattice operations. Instead, the number of conjunctions and disjunctions involved in a sentence determines the dimension of the space of its "meanings": this space is not the fixed dimensional Hilbert space of the states of the system as in QL but a tensorial product of two dimensional Hilbert spaces in which the number of factors varies according to the nature and number of logical connectives. So, the "meaning" of a sentence reflects the logical form of the sentence itself, and this has been pointed as an intensional character of QCL [8]. It also has new connectives that show a genuine quantum behavior, not only when applied to superpositions but also transforming each element of the logical basis in superpositions.

Our work is motivated in two questions posed in [6], [8]. One of them mentions that QCL may be seen as a "new example of fuzzy logics". In this direction, we are interested in the logical formalization of the treatment of quantum information during the computational process. More precisely, we want to establish a lowest bound that allows to appreciate the relevance of inputs that are known with certainty with respect to the possible outputs. The frame we use for this development is the *approximate reasoning*, which is a crucial theme studied within fuzzy logic. In order to accomplish these purposes we must introduce an axiomatization for QCL, what is the second question, referred in [6] as an open problem. This system allows to establish a completeness theorem for the mentioned treatment of information. We first show the relation between QCL and fuzzy logic in a rigorous manner, more precisely its relation with the infinite valued enriched Łukaciewicz calculus. This motivates a way to axiomatization of QCL that reflects the minimum set of basic properties associated to the quantum gates. This axiomatization allows in turn to establish the completeness result mentioned above.

The paper is organized as follows. Section II summarizes basic notions about quantum gates involved in QC. In Section III we present non unitary operations that may be performed on states of physical systems and the Poincaré structure which will be the natural model used as interpretation of the logical system associated with QC. In Section IV we present some preliminary on universal algebra and MV- and PMV-algebras. These structures are the algebraic counterparts of infinite many valued logics. It is necessary here to study injective objects in the variety of PMV-algebras with the purpose to obtain tools for a completeness theorem based on approximate reasoning to be developed later. In Section V we outline the relations between the logic associated to the Poincaré structure and fuzzy logic. In Section VI we introduce an axiomatic system for the irreversible quantum computational structure and show basic results on it. Section VII is dedicated to the study of a fragment of the mentioned logic. From this fragment, in Section VIII we show the formal connection with fuzzy logic and establish a Pavelka style completeness theorem ([13] and [19]) and a type of compactness theorem. Finally, Section IX is devoted to the conclusions.

# 2 BASIC NOTIONS IN THE THEORY OF QUANTUM COMPUTATION

## 2.1 Qbits and qmixes

In quantum computation, information is elaborated and processed by means of quantum systems. Let us first consider the two-dimensional case in which pure states of a system are coherent superpositions of two distinguishable states (may be polarization states of a single photon, spin states of an electron, an so on) represented by unit vectors in the two dimensional complex Hilbert space $\mathcal{C}^2$, and called *qbits*. The standard orthonormal basis $\{|0>, \ |1>\}$ of $\mathcal{C}^2$ is called *the logical basis*. This name refers to the fact that truth is related with $|1>$ and falsity with $|0>$. A qbit $|\varphi>$ may be written as a linear superposition of the basis vectors with complex coefficients (*probability amplitudes*):

$$|\varphi>= c_o|0> +c_1|1>, \quad \text{with} \quad |c_o|^2 + |c_1|^2 = 1$$

Associated with each basis vector is a projector onto the subspace it spans, i.e. $P_o = |0><0|$ and $P_1 = |1><1|$. Due to the usual non complete efficiency of the preparation procedure of physical systems, quantum states are better described by statistical mixtures. They are represented by density operators $\rho$ acting on $\mathcal{C}^2$, called *qmixes*. Qmixes are decomposable in terms of the mentioned projector operators:

$$\rho_\lambda = (1-\lambda)P_o + \lambda P_1, \quad \lambda \text{ a real number } \in \ [0,\ 1]$$

and include pure states as particular cases. We call $\mathcal{D}(\mathcal{C}^2)$ the set of density operators on $\mathcal{C}^2$.

Quantum mechanics reads out the information content of a pure state via the Born rule. By these means, a *probability value* which we will later be interested in may be assigned to a qbit or a qmix in the following way [8]:

**Definition 2.1** Let $|\varphi>= c_o|0> + c_1|1>$ be a qbit, i.e. a unit vector in $\mathcal{C}^2$. Then its *probability value* $p(|\varphi>)$ is $p(|\varphi>) = |c_1|^2$, the square module of the amplitude corresponding to the basis vector associated with truth.

**Definition 2.2** Let $\rho_\lambda = (1-\lambda)P_o + \lambda P_1 \in \mathcal{D}(\mathcal{C}^2)$ be a qmix. Then its *probability value* $p(\rho_\lambda)$ is $p(\rho_\lambda) = \lambda$, the coefficient corresponding to the projector associated with truth.

In order to take into account computations in which multiple qbits are involved, we have to extend the two dimensional Hilbert space to the state space of compound systems. For $n$ qbits, this is performed by taking the n-fold tensor product of state spaces of one qbit. Thus, a unit vector in $\mathcal{C}^2 \otimes \mathcal{C}^2 \otimes \mathcal{C}^2 \otimes \ \ldots \otimes \mathcal{C}^2 \equiv \otimes^n\mathcal{C}^2$ represents the pure state of an *n-qbit* system. We use $x_i$, $1 \leq i \leq n$, as variables ranging over the set $\{0,\ 1\}$ and vectors pertaining to each of the n-logical basis are denoted by $|x_i>$. So, $|x_1> \otimes |x_2> \otimes \ldots \otimes |x_n>$ ( $|x_1, \ \ldots, x_n>$ for short) is a factorized vector of $\otimes^n\mathcal{C}^2$ and $\mathcal{B}^{(n)} = \{|x_1, \ldots, x_n>, x_i \in \{0,\ 1\}\}$ is an orthonormal basis for the n-qbits system, called the *computational basis*. Recall that the dimension of $\otimes^n\mathcal{C}^2$ is $2^n$. Then, pure states of n-qbits systems are normalized linear combinations of the elements of this basis.

To represent qmixes in product spaces, we use the same notation as [8]. So, let us consider the following sets of natural numbers:

$$C_1^{(n)} := \{i : |\ |i>>= |x_1, x_2, \ldots, x_n> \ \text{ with } x_n = 1\}$$

$$C_o^{(n)} := \{i : |\ |i>>= |x_1, x_2, \ldots, x_n> \ \text{ with } x_n = 0\}$$

Any n-qbit $|\varphi> \in \otimes^n\mathcal{C}^2$

$$|\varphi>= \sum_{i=0}^{2^n-1} a_i|\ |i>>$$

may be expressed as

$$|\varphi >= \sum_{i \in C_o^{(n)}} a_i|\ |i >> + \sum_{j \in C_1^{(n)}} |\ |j >>$$

$P_o^{(n)}$ and $P_1^{(n)}$ are the projectors onto the subspaces spanned by $\{|\ |i >>$: $i \in C_o^{(n)}\}$ and $\{|\ |i >>$: $i \in C_1^{(n)}\}$ respectively. The corresponding density operators associated to them are $k_n P_o^{(n)}$ and $k_n P_1^{(n)}$, with

$$k_n = \frac{1}{2^n - 1}.$$

Let now $\mathcal{D}(\otimes^n \mathcal{C}^2)$ be the set of all density operators of $\otimes^n \mathcal{C}^2$ and

$$\mathcal{D} := \bigcup_{n=1}^{\infty} \mathcal{D}(\otimes^n \mathcal{C}^2)$$

Then,

**Definition 2.3** A qmix is a density operator in $\mathcal{D}$.

The probability value of an n-qmix is also given by the Born rule:

**Definition 2.4** For any $\rho \in \mathcal{D}$, the *probability value* of $\rho$ is

$$p(\rho) = Tr(P_1^{(n)} \rho)$$

The quantum computation process may be summarized as follows: first an initial state of a physical system (a qmix) is provided as the *input*. Then, it evolves according to the *elementary operations* that are performed on it, each one associated to a quantum gate. Finally, the access to the information content of the resulting state is possible via the measurement operation that provides one of the possible results, more about this below when we deal with approximate reasoning.

## 2.2 Quantum gates on qbits

In QCL, logical connectives are interpreted as transition operations on the physical system and represented by quantum gates. Quantum gates are implemented as processes that transform the state of the system at a certain initial time $t_o$ to its state at a later final time $t$. Due to the Schrödinger equation, this change is given by an unitary operator $U(t_o,\ t)$ if the evolution is not explicitly depending on time. So,

$$\rho_\lambda(t) = U(t_o,\ t)\rho_\lambda(t_o)U^\dagger(t_o,\ t)$$

where $U^\dagger$ is the adjoint of $U$. Unitary operations preserve vector lengths and angles. Moreover, they are reversible operations, so the process taking from $t_o \longrightarrow t$ may be reversed without any loose of information, as Charles Bennett first announced in the early '70's [3], exception made of the reading of the result or the ignoring of some degrees of freedom, to what we will turn on later. We need to introduce here the following quantum gates acting on qbits:

**Definition 2.5** The **noop** gate applied on a qbit is the operation that does nothing.

The **noop** gate is the identity operation. We will call it **I**. It is immediately generalized to the case of n-qbits.

**Definition 2.6** The **not** (*negation*) gate applied on a qbit is the linear operation **not**: $\mathcal{C}^2 \to \mathcal{C}^2$ that acts over the logical basis in the following way:

$$\mathbf{not}|0>=|1>$$

$$\mathbf{not}|1>=|0>$$

Its corresponding matrix representation in the logical basis is

$$\begin{pmatrix} 0 & 1 \\ 1 & 0 \end{pmatrix}$$

i.e. the $\sigma_x$ Pauli matrix (see Section 3).

**Definition 2.7** The $\mathbf{not}^{(j)}$ (*negation of the j-register*) gate applied on an n-qbit is the linear operation $\mathbf{not}^{(j)}$: $\otimes^n\mathcal{C}^2 \to \otimes^n\mathcal{C}^2$ that acts over any element of the computational basis in the following way:

$$\mathbf{not}^{(j)}|x_1, ..., x_n>=|x_1, ..., x_{j-1},\ 1 - x_j, x_{j+1}, ..., x_n>$$

Its corresponding $2^n \times 2^n$ matrix representation in the computational basis is the block matrix with identities in the diagonal everywhere except for the block acting on $x_j$, where **I** is replaced by the $\sigma_x$ matrix Pauli.

**Definition 2.8** The $\sqrt{\mathbf{not}}$ (*square root of the negation*) gate applied on a qbit is the linear operator $\sqrt{\text{not}} : \mathcal{C}^2 \to \mathcal{C}^2$ that acts over the logical basis in the following way:

$$\sqrt{\mathbf{not}}\ |0> = \frac{1+\imath}{2}|0> + \frac{1-\imath}{2}|1>$$

$$\sqrt{\mathbf{not}}\ |1> = \frac{1-\imath}{2}\ |0> + \frac{1+\imath}{2}\ |1>$$

The corresponding matrix representation in the logical basis is

$$\begin{pmatrix} \frac{1+\imath}{2} & \frac{1-\imath}{2} \\ \frac{1-\imath}{2} & \frac{1+\imath}{2} \end{pmatrix}$$

Physically, it is a device that, whenever the input is, for example, $|0>$ the output is

$$\begin{cases} |0> & \text{with probability } \ 1/2 \ , \\ |1> & \text{also with probability } \ 1/2 \end{cases}$$

and analogously for $|1>$. So it looks like a random switch. It may be thought that, applying it twice, it would go on behaving the same way. But this is not the case. Instead, applying it twice gives the usual negation **not**. This is precisely the property that characterizes $\sqrt{\mathbf{not}}$:

$$\sqrt{\mathbf{not}}\sqrt{\mathbf{not}}|\varphi> = \mathbf{not}|\varphi>$$

This is due to the fact that, when evaluating the probability of a quantum transition, we have to add the probability amplitudes of all possible paths and not the probabilities themselves. Probability amplitudes are complex numbers that may cancel (destructive interference) or enhance each other (constructive interference) when summed up and this makes a big difference with classical probabilistic processes (see, for example, [9], [1]). The potential of quantum computing lies precisely in quantum interference among many computational paths. $\sqrt{\mathbf{not}}$ is sometimes interpreted as a "tentative" negation: trying twice effectively produces negation [8].

When applied to an n-qbit we have:

**Definition 2.9** The $\sqrt{\text{not}}^{(j)}$ gate applied on an n-qbit is the linear operator $\sqrt{\text{not}}^{(j)} : \otimes^n\mathcal{C}^2 \to \otimes^n\mathcal{C}^2$ that acts on any $|\varphi>$ in the following way:

$$\sqrt{\text{not}}^{(j)}\ |\varphi> = \sum_{i=0}^{2^n-1} a_i|x_1, x_2, \ldots, x_{j-1}, x_{j+1}, \ldots, x_n> \ \otimes\ \sqrt{\text{not}}\ |x_j>$$

Its matrix representation is constructed analogously to the **not** gate.

To consider the (reversible) *conjunction*, the Toffoli gate is needed. Let us suppose that two multiple qbits $|\varphi> \in \otimes^n \mathcal{C}^2$ and $|\phi> \in \otimes^m \mathcal{C}^2$ are given. Denoting by $\otimes^{n+m+1}\mathcal{C}^2 \equiv (\otimes^n \mathcal{C}^2) \otimes (\otimes^m \mathcal{C}^2 \otimes \mathcal{C}^2)$,

**Definition 2.10** The *Toffoli* gate $T^{(n,m,1)}$ is the linear operator

$$T^{(n,m,1)} : \otimes^{n+m+1}\mathcal{C}^2 \to \otimes^{n+m+1}\mathcal{C}^2$$

acting over any vector $|x_1, x_2, ..., x_n > \otimes |y_1, y_2, ..., y_m > \otimes |z >$ in the computational basis in the following way:

$$T^{(n,m,1)}|x_1, x_2, ..., x_n > \otimes |y_1, y_2, ..., y_m > \otimes |z >=$$

$$|x_1, x_2, ..., x_n > \otimes |y_1, y_2, ..., y_m > \otimes |min(x_n, y_m) \oplus z >$$

where $\oplus$ indicates the sum modulo 2.

On this basis, the (reversible) *conjunction* **and** may be defined.

**Definition 2.11** For any $|\varphi> \in \otimes^n \mathcal{C}^2$ and $|\phi> \in \otimes^m \mathcal{C}^2$,

$$\mathbf{and}(|\varphi>, |\phi>) := T^{(n,m,1)}|\varphi> \otimes |\phi> \otimes |0>$$

We will deal little with **and** in what follows. We only remark here that the explicit writing in the computational basis of the n-qbits representing the pieces of information related by **and** reproduces the results in the truth table of the conjunction, added each file with the probability amplitude whose square modulus represents the corresponding probability of obtaining that result if measured. For considerations about this connective we refer to [6].

## 2.3 Quantum gates on qmixes

These gates are easily generalized to qmixes. As in [8], we use capital letters for gates acting on qmixes.

**Definition 2.12** The **NOOP** gate applied on an n-qmix $\rho \in \mathcal{D}$ does nothing

**Definition 2.13** The **NOT** (*negation*) gate applied on a qmix $\rho_\lambda \in \mathcal{D}(\mathcal{C}^2)$ gives

$$\mathbf{NOT}\ \rho_\lambda = \mathbf{not}\rho_\lambda\mathbf{not}$$

As expected, **NOT** gate satisfies that

**Proposition 2.14** *Let $\lambda$ be a real number in $[0,1]$. If we consider the density operator $\rho_\lambda = (1-\lambda)P_0 + \lambda P_1 \in \mathcal{D}(\mathcal{C}^2)$ then we have*

$$p(\mathbf{NOT}\rho_\lambda) = 1 - \lambda$$

**Definition 2.15** The $\mathbf{NOT}^{(n)}$ gate applied on a density operator $\rho \in \mathcal{D}$ gives

$$\mathbf{NOT}^{(n)}\ \rho = \mathbf{not}^{(n)} \rho \mathbf{not}^{(n)}$$

**Definition 2.16** The $\sqrt{\mathbf{NOT}}$ (*square root of the negation*) gate applied on a qmix $\rho_\lambda \in \mathcal{D}(\mathcal{C}^2)$ gives

$$\sqrt{\mathbf{NOT}}\ \rho_\lambda = \sqrt{\mathbf{not}}\ \rho_\lambda\ \sqrt{\mathbf{not}}^\dagger$$

where $^\dagger$ means the adjoint operation.

The previously discussed behavior of square root of the negation on qbits reflects for qmixes in the fact that

**Proposition 2.17** *Let $\lambda$ be a real number in $[0,1]$. If we consider the density operator $\rho_\lambda = (1-\lambda)P_0 + \lambda P_1 \in \mathcal{D}(\mathcal{C}^2)$ then we have*

$$p(\sqrt{\mathbf{NOT}}\rho_\lambda) = \frac{1}{2}$$

*Proof:* See [8] □

**Definition 2.18** The $\sqrt{\mathbf{NOT}}^{(n)}$ gate applied on a density operator $\rho \in \mathcal{D}$ gives

$$\sqrt{\mathbf{NOT}}^{(n)}\ \rho = \sqrt{\mathbf{not}}^{(n)}\ \rho\ \sqrt{\mathbf{not}}^{(n)\dagger}$$

The (reversible) *conjunction* **AND** of q-mixes $\in \mathcal{D}(\mathcal{C}^2)$ is defined as

**Definition 2.19** Consider $\tau_\lambda \in \mathcal{D}(\mathcal{C}^2)$ and $\nu_\gamma \in \mathcal{D}(\mathcal{C}^2)$, then

$$\mathbf{AND}(\tau_\lambda, \nu_\gamma) = T^{(1,1,1)}(\tau_\lambda \otimes \nu_\gamma \otimes P_o^{(1)})T^{(1,1,1)}$$

and in the case of greater dimensions,

**Definition 2.20** Consider $\tau \in \mathcal{D}(\otimes^n \mathcal{C}^2)$ and $\nu \in \mathcal{D}(\otimes^m \mathcal{C}^2)$, then

$$\mathbf{AND}(\tau, \nu) = T^{(n,m,1)}(\tau \otimes \nu \otimes P_o^{(1)})T^{(n,m,1)}$$

In order to access information after the computing process we apply the *measurement device* **meas**. It is not a unitary operation but an intrinsically probabilistic process that, when applied to a qbit, converts it into a probabilistic classical bit that is 0 (falsity) or 1 (truth) associated with the elements of the logical basis $\{|0>, |1>\}$ with probability equal to the modulus squared of its amplitude, i.e.

**Definition 2.21** Given $|\varphi> = c_o|0> + c_1|1>$, **meas** device over it produces

$$\begin{cases} 0, \ |0> & \text{with probability } \ |c_o|^2 \ , \\ 1, \ |1> & \text{with probability } \ |c_1|^2 \end{cases}$$

In the cases of qmixes $\rho_\lambda \in \mathcal{D}(\mathcal{C}^2)$, the results are:

$$\begin{cases} 0, P_o & \text{with probability (1-}\lambda\text{ ),} \\ 1, \ P_1 & \text{with probability } \lambda \end{cases}$$

In spite of the fact that quantum gates are unitary operations, quantum algorithms are not deterministic due to the randomness of the measurement process. Instead, there is a probability distribution of the possible outcomes. Nevertheless, they are useful because the output may be the correct solution to a very hard problem but such that the solution could be easily verified, as is the case of finding the prime factors of a large number -a hard problem whose solution is easily verifiable.

In what follows, we will deal with states represented by qmixes because they are the general case and include pure states.

# 3 POINCARE IRREVERSIBLE QUANTUM STRUCTURE

First we recall some properties of qbits and qmixes that we will use in what follows. We restrict first to $\mathcal{C}^2$. Let us consider the Bloch sphere, i.e. the sphere $S^2$ with unit radius:

$$S^2 = \{x_i \in \mathcal{R}^3 / \sum_{i=1}^{3} x_i^2 = 1\}$$

The coefficients $c_o$ and $c_1$ of the decomposition in the logical basis of any qbit $|\varphi>$ may be written as

$$c_o = \cos\frac{\theta}{2} \qquad c_1 = \exp(\imath\phi)\sin\frac{\theta}{2}$$

i.e. in terms of the Euler angles that determine a point of the Bloch sphere. Thus, a qbit may be seen as a point on the surface of the Bloch sphere. The transformation of the qbit due to the action of a gate in the computational process is a unitary transformation in $\mathcal{C}^2$, i.e. an element of $SU(2)$. So, because of the relation $SO(3) \cong SU(2)/Z_2$, a rotation in $\mathcal{R}^3$ corresponds to it. Thus, any transformation applied to the qbit may be considered a rotation.

Qmixes and any operator over $\mathcal{C}^2$ have a matrix representation

$$A = aI + a_x\sigma_x + a_y\sigma_y + a_z\sigma_z$$

where $\sigma_x$, $\sigma_y$, $\sigma_z$ are the Pauli matrices. In the logical basis:

$$\sigma_x = \begin{pmatrix} 0 & 1 \\ 1 & 0 \end{pmatrix} \qquad \sigma_y = \begin{pmatrix} 0 & -i \\ i & 0 \end{pmatrix} \qquad \sigma_z = \begin{pmatrix} 1 & 0 \\ 0 & -1 \end{pmatrix}$$

In particular, because of the properties of density operators $tr\rho = 1$, $\rho^\dagger = \rho$, the matrix representation of $\rho$ may be written as

$$\rho = \frac{1}{2}(I + r_1\sigma_x + r_2\sigma_y + r_3\sigma_z)$$

where $r_1$, $r_2$, $r_3$ are real numbers such that $r_1^2 + r_2^2 + r_3^2 \leq 1$. Thus each density operator $\rho$ in $\mathcal{D}(\mathbf{C}^2)$ such that $\rho^2 = \rho$ (pure states) can be identified to a point $(r_1, r_2, r_3)$ on the sphere and each $\rho$ in $\mathcal{D}(\mathcal{C}^2)$ such that $\rho^2 < \rho$ (proper mixtures) with a point in the interior of the Bloch sphere. We denote this identifications as $\rho = (r_1, r_2, r_3)$.

**Remark 3.1** We recall that the probability value of a qmix $\rho$ in $\mathcal{D}(\mathcal{C}^2)$ is a function $p : \mathcal{D}(\mathcal{C}^2) \to [0, 1]$ determined by the Born rule $p(\rho) = Tr(\rho P_1)$. When expanding $\rho$ in terms of the Pauli matrices, it results that it depends only on the $r_3$ coefficient. So, the semantic consequence will be only associated with $r_3$. Moreover, it may be easily seen that the probability value of $p(\sqrt{\mathbf{NOT}}\ \rho)$ is only related to the $r_2$ coefficient of the expansion.

Precisely,

**Lemma 3.2** [8, Lemma 6.1]) *Let* $\rho = (r_1, r_2, r_3) \in \mathcal{D}(\mathcal{C}^2)$. *Then we have*

1. $p(\rho) = \frac{1-r_3}{2}$,
2. $p(\sqrt{\mathbf{NOT}}\ \rho) = \frac{1-r_2}{2}$.

□

Up to now we have been dealing with reversible gates representing unitary evolutions. But there are other interesting operations that may be done with qbits or qmixes. Namely, ignoring some degrees of freedom of the system and paying attention only to the others, what is performed by taking partial traces over the density operator of the system. This is, the same as **meas**, an irreversible operation and, in the case of taking a partial trace over pure states, this operation transforms it in a mixture.

Let us consider the state space $(\otimes^n \mathcal{C}^2) \otimes (\otimes^m \mathcal{C}^2) \otimes (\otimes^r \mathcal{C}^2)$. For any state $\rho$, there is a unique density operator $\tilde{\rho} = Tr_{1,2}(\rho)$, the partial trace of $\rho$, called its *reduce state*, that represents the state obtained by ignoring the information related to the degrees of freedom in $\otimes^n \mathcal{C}^2$ and in $\otimes^m \mathcal{C}^2$. The states $\rho$ and $\tilde{\rho}$ are statistically equivalent with respect to the subsystem whose state space is $\otimes^r \mathcal{C}^2$, in the sense that the mean value $< O >$ of an observable $O$ associated with this subsystem may be calculated by using any of them. That is: let $O : \otimes^r \mathcal{C}^2 \to \otimes^r \mathcal{C}^2$ and let $\widetilde{O} = I^{(n)} \otimes I^{(m)} \otimes O : (\otimes^n \mathcal{C}^2) \otimes (\otimes^m \mathcal{C}^2) \otimes (\otimes^r \mathcal{C}^2) \to (\otimes^n \mathcal{C}^2) \otimes (\otimes^m \mathcal{C}^2) \otimes (\otimes^r \mathcal{C}^2)$ be its extension to the complete space state, then $< O >= Tr(\tilde{\rho} O) = Tr(\rho \widetilde{O})$. This consideration of the reduced state allows the definition of another conjunction, besides the reversible conjunction represented by the Toffoli gate, called *irreversible conjunction* **IAND**.

**Definition 3.3** Let $\tau$ and $\nu$ in $\mathcal{D}(\mathcal{C}^2)$. Then the *irreversible conjunction* is defined as

$$\mathbf{IAND}(\tau,\ \nu) = \rho_{p(\tau)p(\nu)}$$

It may be seen that this operation is equivalent to a partial trace over the state $\mathbf{AND}(\tau,\ \nu)$ representing the reversible conjunction of the states $\tau$, $\nu$ of two systems ([8]):

**Proposition 3.4** *Consider the reversible conjunction*

$$\mathbf{AND}(\tau, \nu) = T^{(1,1,1)}(\tau, \nu, P_o) T^{(1,1,1)} \in \mathcal{C}^2 \otimes \mathcal{C}^2 \otimes \mathcal{C}^2$$

*then* $\mathbf{IAND}(\tau, \nu) = Tr_{1,2}(\mathbf{AND}(\tau, \nu))$ □

Another connective may be defined associated with an irreversible operation, namely the *Lukasiewicz disjunction.*

**Definition 3.5** Let $\tau$ and $\nu$ in $\mathcal{D}(\mathcal{C}^2)$. Then, the *Lukasiewicz disjunction* is defined as

$$\tau \oplus \nu = \rho_{p(\tau) \oplus p(\nu)}$$

From now on and for simplicity in the logic algebraic handling, we will use the following conventions: $\neg\rho$ is the abbreviation for $(\mathbf{NOT}\ \rho)$, $\sqrt{\rho}$ is the abbreviation for $(\sqrt{\mathbf{NOT}}\ \rho)$ and $(\rho \bullet \sigma)$ is the abbreviation for $\mathbf{IAND}(\rho, \sigma)$.

Taking into account these conventions, we introduce the following algebraic system associated with the quantum gates:

**Definition 3.6** The *Poincaré irreversible quantum computational algebra* (for short: Poincaré IQC-algebra) is given by

$$\langle \mathcal{D}(\mathbf{C}^2), \bullet, \oplus, \neg, \sqrt{,} P_o, P_1, \rho_{1/2} \rangle$$

$P_o, P_1$ and $\rho_{1/2}$ are distinguished elements of $\mathcal{D}(\mathbf{C}^2)$ representing privileged states associated with *true, false* and *random* respectively.
Further operation are defined as follows:

$$\alpha \odot \beta = \neg(\neg\alpha \oplus \neg\beta)$$

$$\alpha \to \beta = \neg\alpha \oplus \beta$$

$$\alpha \wedge \beta = \alpha \odot (\varphi \to \beta)$$

$$\alpha \vee \beta = (\alpha \to \beta) \to \beta$$

These operations satisfy the following properties:

**Lemma 3.7** *Let $\tau, \nu \in \mathcal{D}(\mathbf{C}^2)$. Then we have:*

1. $\langle \mathcal{D}(\mathbf{C}^2), \bullet \rangle$ *is an abelian monoid,*
2. $\tau \bullet P_o = P_o$
3. $\tau \bullet P_1 = \rho_{p(\tau)}$

4. $p(\tau \bullet \nu) = p(\tau)p(\nu)$

5. $p(\tau \oplus \nu) = p(\tau) \oplus p(\nu)$

6. $\sqrt{\neg\tau} = \neg\sqrt{\tau}$

7. $\sqrt{\sqrt{\tau}} = \neg\tau$

8. $p(\sqrt{\tau \bullet \nu}) = p(\sqrt{\tau \oplus \nu}) = 1/2$

9. $\frac{p(\sigma)}{4} \oplus \frac{p(\sqrt{\sigma})}{4} \leq \frac{1+\sqrt{2}}{4\sqrt{2}}$ if and only if $r_2^2 + r_3^2 \leq 1$ *with* $\sigma = \sigma(r_1, r_2, r_3)$

10. $\frac{p(\sigma)}{4} \oplus \frac{1}{8} \leq \frac{3}{8} \leq \frac{1+\sqrt{2}}{4\sqrt{2}}$.

*Proof:* For 1...8 see [8, Lemma 6.2 and Lemma 5.3]. 9) We use the factor 1/4 since for each $x, y \in [0,1]$, $1/4x \oplus 1/4y = 1/4x + 1/4y$. By Proposition 3.2, $\frac{p(\sigma)}{4} + \frac{p(\sqrt{\sigma})}{4} = (1 - r_3)/8 + (1 - r_2)/8$. If we suppose that $(r_2, r_3)$ lies in the unitary circle then this bound is given by simple problem of maximization with conditional extremes. To see the converse, taking into account Proposition 3.2, $(1-r_3)/2 + (1-r_2)/2 \leq 1 + 1/\sqrt{2}$, thus $-(r_2 + r_3) \leq 2/\sqrt{2}$. Using polar coordinates $r_2 = r\cos\theta$, $r_3 = r\sin\theta$, thus we have $-r(\cos\theta + \sin\theta) \leq 2/\sqrt{2}$. But the maximum of $-r(\cos\theta + \sin\theta)$ is given when $\theta = 5/4\pi$, in this case $(\cos\theta + \sin\theta) = -2/\sqrt{2}$ and the $r \leq 1$. Thus $(r_2, r_3)$ lies in the unitary circle. 10) immediate. □

# 4 PRELIMINARIES ON UNIVERSAL ALGEBRA. MV-ALGEBRAS AND PMV-ALGEBRAS

In this section we first deal with basic notions on universal algebra that we will apply to the PMV structures which are the algebraic counterpart of a type of enriched Łukaciewcz infinite valued calculus. This calculus will play a fundamental role in the logic associated with quantum gates. We recall from from [2] and [5] some basic tools of injectives and universal algebra respectively. Let $\mathcal{A}$ be a class of algebras. For all algebras $A, B$ in $\mathcal{A}$, $[A, B]_{\mathcal{A}}$ will denote the set of all homomorphism $g : A \to B$. The class $\mathcal{A}$ is said to be a variety if and only if it is definable by equations or, equivalently, it is closed by sub algebras, direct product and homomorphic images. An algebra $A$ in $\mathcal{A}$ is *injective* if and only if for every $f \in [B, A]_{\mathcal{A}}$ and every monomorphism $g \in [B, C]_{\mathcal{A}}$ there exists $h \in [C, A]_{\mathcal{A}}$ such that the following diagram is commutative

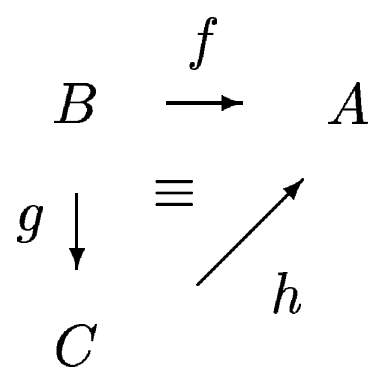

$A$ is *self-injective* if and only if every homomorphism from a subalgebra of $A$ into $A$, extends to an endomorphism of $A$. For each algebra $A$, we denote by $Con(A)$, the congruence lattice of $A$, the diagonal congruence is denoted by $\Delta$ and the largest congruence $A^2$ is denoted by $\nabla$. An algebra $I$ is *simple* if and only if $Con(I) = \{\Delta, \nabla\}$. An algebra $A$ is *semisimple* if and only if it is a subdirect product of simple algebras. A nontrivial algebra $T$ is said to be *minimal* in $\mathcal{A}$ if and only if for each nontrivial algebra $A$ in $\mathcal{A}$, there exists a monomorphism $f : T \to A$. A simple algebra $I_M$ is said to be *maximum simple* if and only if for each simple algebra $I$, $I$ can be embedded in $I_M$. A simple algebra is *hereditarily simple* if and only if all its subalgebras are simple. An algebra $A$ is *rigid* if and only if the identity homomorphism is the only automorphism. An algebra $A$ has the *congruence extension property* (CEP) if and only if for each subalgebra $B$ and $\theta \in Con(B)$ there is a $\phi \in Con(A)$ such that $\theta = \phi \cap A^2$. A variety $\mathcal{V}$ satisfies CEP if and only if every algebra in $\mathcal{V}$ has the CEP. It is clear that if $\mathcal{V}$ satisfies CEP then every simple algebra is hereditarily simple.

**Definition 4.1** Let $\mathcal{V}$ be a variety. Two constant terms $0, 1$ of the language of $\mathcal{V}$ are called *distinguished constants* if and only if $A \models 0 \neq 1$ for each nontrivial algebra $A$ in $\mathcal{V}$.

**Theorem 4.2** *Let $\mathcal{A}$ be a variety satisfying CEP, with distinguished constants* $0, 1$. *If $I$ is a self-injective maximum simple algebra in $\mathcal{A}$ then $I$ is injective.*

*Proof:* ([11, Theorem 3.4]). □

**Definition 4.3** *An MV-algebra* [7] *is an algebra* $\langle A, \oplus, \neg, 0\rangle$ *of type* $\langle 2, 2, 0\rangle$ *satisfying the following equation*

*MV1* $\langle A, \oplus, 0\rangle$ *is an abelian monoid,*

*MV2* $\neg\neg x = x$

*MV3* $x \oplus \neg 0 = \neg 0$

*MV4* $\neg(\neg x \oplus y) \oplus y = \neg(\neg y \oplus x) \oplus x$

We denote by $\mathcal{MV}$ the variety of MV-algebras. In agreement with the usual MV-algebraic operations we define

$$1 = \neg 0$$

$$x \odot y = \neg(\neg x \oplus \neg y)$$

$$x \rightarrow y = \neg x \oplus y$$

$$x \wedge y = x \odot (x \rightarrow y)$$

$$x \vee y = (x \rightarrow y) \rightarrow y$$

On each MV-algebra $A$ we can define an order in $A$ given by

$$x \leq y \text{ if and only if } x \rightarrow y = 1$$

This order turns $\langle A, \wedge, \vee, 0, 1\rangle$ into a distributive bounded lattice with 1 the greatest element and 0 the smallest element.

A very important example of MV-algebra is $[0,1]_{MV} = \{[0,1], \oplus, \neg, 0\}$ such that $[0,1]$ is the real unit segment and $\oplus$ and $\neg$ are defined as follows

$$x \oplus y = min(1, x + y)$$

$$\neg x = 1 - x$$

The derivate operations in $[0,1]_{MV}$ are given by $x \odot y = max(0, x + y - 1)$ and $x \rightarrow y = min(1, 1 - x + y)$. Finally the MV-lattice structure is the natural order in $[0,1]$.

We recall from [13] some well-known facts about implicative filters and congruences on MV-algebras. To do this, we will see MV-algebras as particular case of BL-algebras.

Let A be an MV-algebra and $F \subseteq A$. Then $F$ is an *implicative filter* if and only if it satisfies the following conditions:

1. $1 \in F$,
2. if $x \in F$ and $x \rightarrow y \in F$ then $y \in F$.

It is easy to verify that a nonempty subset $F$ of a MV-algebra $A$ is an implicative filter if and only if for all $a, b \in A$:

- If $a \in F$ and $a \leq b$ then $b \in F$,

- if $a, b \in F$ then $a \odot b \in F$.

Note that an implicative filter $F$ is proper if and only if 0 does not belong to $F$. The intersection of any family of implicative filters of $A$ is again an implicative filter of $A$. We denote by $\langle X \rangle$ the implicative filter generated by $X \subseteq A$, i.e., the intersection of all implicative filters of $A$ containing $X$. We abbreviate this as $\langle a \rangle$ when $X = \{a\}$ and it is easy to verify that $\langle X \rangle = \{x \in A : \exists\ w_1 \cdots w_n \in X \text{ such that } x \geq w_1 \odot \cdots \odot w_n\}$. For any implicative filter $F$ of $A$, $\theta_F = \{(x, y) \in A^2 : x \to y, y \to x \in F\}$ is a congruence on $A$. Moreover $F = \{x \in A : (x, 1) \in \theta_F\}$. Conversely, if $\theta \in Con(A)$ then $F_\theta = \{x \in A : (x, 1) \in \theta\}$ is an implicative filter and $(x, y) \in \theta$ if and only if $(x \to y, 1) \in \theta$ and $(y \to x, 1) \in \theta$. Thus the correspondence $F \to \theta_F$ is a bijection from the set of implicative filters of $A$ onto the set $Con(A)$. If $F$ is an implicative filter of $A$, we shall write $A/F$ instead of $A/\theta_F$, and for each $x \in A$ we shall write $x/\theta_F$ for the equivalence class of $x$. An implicative filter is called *prime* if and only if, for each $x, y \in A$, $x \to y \in F$ or $y \to x \in F$.

**Proposition 4.4** *Let $A$ be an MV-algebra and $F$ be an implicative filter in $A$. Then we have: $A/F$ is tottaly ordered if and only if $F$ is prime*

*Proof:* See [13, Lemma 2.3.14]. □

**Theorem 4.5** *For every MV-algebra $A$ the following conditions are equivalent:*

1. *$A$ is simple,*

2. *$A$ is no trivial and for every nonzero element $x \in A$, there is an integer $n > 0$ such that $1 = x \oplus \cdots \oplus x$ ($n$ times),*

3. *$A$ is no trivial and for each element $x < 1$ in $A$, there is an integer $n > 0$ such that $0 = x \odot \cdots \odot x$ ($n$ times),*

4. *$A$ is isomorphic a subalgebra of $[0, 1]_{MV}$.*

*Proof:* See [7, Theorem 3.5.1]. □

**Proposition 4.6** *Two subalgebras $A$, $B$ of $[0,1]_{MV}$ are isomorphic if and only if $A = B$; the identity function is the only automorphism of $A$*

*Proof:* See [7, Corollary 7.2.6] $\square$

**Corollary 4.7** *1.* $[0,1]_{MV}$ *is the maximum simple algebra in* $\mathcal{MV}$.

*2.* $[0,1]_{MV}$ *is hereditary simple algebra.*

*3.* $[0,1]_{MV}$ *is self injective.*

$\square$

**Definition 4.8** *A product MV-algebra* [16] *(for short: PMV-algebra) is an algebra* $\langle A, \oplus, \bullet, \neg, 0\rangle$ *of type* $\langle 2,2,1,0\rangle$ *satisfying the following*

*1* $\langle A, \oplus, \neg, 0\rangle$ *is an MV-algebra*

*2* $\langle A, \bullet, 1\rangle$ *is an abelian monoid*

*3* $x \bullet (y \odot \neg z) = (x \bullet y) \odot \neg(x \bullet z)$

We denote by $\mathcal{PMV}$ the variety of PMV-algebras. An important example of PMV-algebra is $[0,1]_{MV}$ equipped with multiplication. This algebra is denoted by $[0,1]_{PMV}$. Note that every Boolean algebra becomes a PMV-algebra by letting the product operation coincide with the infimun operation. The following are almost immediate consequences of the definition.

**Lemma 4.9** *In each PMV-algebra we have*

*1.* $0 \bullet x = 0$

*2. If* $a \leq b$ *then* $a \bullet x \leq b \bullet x$

*3.* $x \odot y \leq x \bullet y \leq x \wedge y$

$\square$

**Definition 4.10** *If we consider* $1/2 \in [0,1]$ *then the algebra* $S$ *is the sub algebra of* $[0,1]_{PMV}$ *generated by* $1/2$

It is clear that, as a set, $S$ is contained in the rationals of the real interval $[0,1]$. The algebra $S$ and the following proposition will play an important role in the later logical treatment of irreversible quantum gates.

**Proposition 4.11** *$(S, \leq)$ is dense order in the real interval* $[0, 1]$.

*Proof:* Let $a, b \in [0, 1]$. We assume that $a < b < 1/2$. Let $n_0 \in N$ the first natural such that $1/2^{n_0} \leq a$. Let $n_1 \in N$ such that $1/2^{n_1} \leq 1/4\ min\{b - a,\ a - 1/2^{n_0}\}$. Thus there exists $n \in N$ such that

$$a < s = 1/2^{n_0} + \sum_n 1/2^{n_1} = 1/2^{n_0} + \bigoplus_n 1/2^{n_1} < b$$

and, by construction, $s \in S$. If $1/2a < b$, then $\neg b < \neg a < 1/2$ since $\neg 1/2 = 1 - 1/2$. Therefore, we find $s \in S$ such $\neg b < s < \neg a$ following the same argument. Thus $a < \neg s = 1 - s < b$. Finally $(S, \leq)$ is a dense order in the real interval $[0, 1]$. □

**Lemma 4.12** *Let $A$ be a PMV-algebra and let $B$ be the underlying MV-algebra. Then $A$ and $B$ have the same congruences.*

*Proof:* ([16, Lemma 2.11]). □

In view of this, implicative filters on a PMV-algebra $A$ are identifiable to congruences of $A$, thus we use the notation $A/F$ for the quotient PMV-algebra.

**Corollary 4.13** *Let $A$ be a PMV-algebra and $F$ be an implicative filter. Then we have: $A/F$ is totally ordered if and only if $F$ is a prime implicative filter.*

*Proof:* Follows from Lemma 4.4 □

**Corollary 4.14** *If $A$ is a subalgebra of $[0, 1]_{PMV}$, then $A$ is simple.*

*Proof:* Follows from Theorem 4.5 Lemma 4.12 □

**Theorem 4.15** *Let $A$ be a semisimple MV-algebra. There is at most one operation $\bullet$ making $(M, \bullet)$ into a PMV-algebra.*

*Proof:* See [17, Theorem 3.1.14]. □

**Proposition 4.16** *If $\mathcal{A}$ is a subvariety of $\mathcal{PMV}$, then $\mathcal{A}$ satisfies CEP.*

*Proof:* Let $A$ be a PMV-algebra and let $B$ be a sub algebra of $A$. By Lemma 4.12, our proof is based only in the notion of implicative filters. For each implicative filter $F$ of $B$, let $\langle F\rangle_A$ be the implicative filter of $A$ generated by $F$. Clearly $F \subseteq \langle F\rangle_A$. To see the converse, let $b \in B \cap \langle F\rangle_A$. Then there exists $a_1, \cdots, a_n \in F$ such that $a_1 \odot a_2 \odot \cdots \odot a_n \leq b$. Since $b \in B$ and $F$ is an implicative filter of B, hence upward closed, it follows that $b \in F$. □

**Theorem 4.17** *$[0,1]_{PMV}$ is injective in $\mathcal{PMV}$.*

*Proof:* It is clear that $0, 1$ are distinguished constant terms in $\mathcal{PMV}$. By Proposition 4.16, $\mathcal{PMV}$ satisfies CEP. We want to see that $[0,1]_{PMV}$ is a maximum simple algebra in $\mathcal{PMV}$. Let $A$ be a simple algebra in $\mathcal{PMV}$. By Lemma 4.12, $A$ is simple as MV-algebra. Thus, by Corollary 4.7.1 there exists a sub MV-algebra $A_0$ of $[0,1]_{MV}$ isomorphic to $A$. Let $i : A \rightarrow A_0$ an MV-isomorphism, then $i$ induces an operation $\bullet$ on $A_0$ making $(A_0, \bullet)$ into a PMV-algebra. More precisely: If $x, y \in A_0$ such that $x = i(a)$ and $y = i(b)$ then $x \bullet y = i(a \bullet b)$. Since $A_0$ is a simple MV-algebra (Corollary 4.7.3), by Theorem 4.15, the only possible candidate to be the $\bullet$ such that $(A_0, \bullet)$ turns to be a PMV-algebra is the usual product inherited from $[0,1]$, because it is an abelian monoid with 1 as neutral element and satisfies $x(y \odot \neg z) = (xy) \odot \neg(xz)$. Thus, for each $a, b \in A$, $i(a \bullet b) = i(a)i(b)$ resulting $i : A \rightarrow A_0 \subseteq [0,1]_{PMV}$ a PMV-monomorphism. This proves that $[0,1]_{PMV}$ is a maximum simple algebra. Using the last argument and Corollary 4.7.3, $[0,1]_{PMV}$ is self injective. Finally we are within the hypothesis of Theorem 4.2, resulting $[0,1]_{PMV}$ injective in $\mathcal{PMV}$. □

# 5 MOTIVATION OF THE LOGICAL CALCULUS ASSOCIATED WITH $IQC$-ALGEBRA

In [8] it is said that "the logic arising from quantum computation represents, in a sense, a new example of fuzzy logic" and also that its axiomatization is an open problem. In this section, we begin to rigorously delineate the relation between the logic associated to the Poincaré structure and fuzzy logic from a semantic approach.

This will be done by establishing a logical propositional system in which propositions are interpreted in $\mathcal{D}(\mathcal{C}^2)$ in a way that the semantic consequence $\models$ comes from the relation $p(\sigma) \leq p(\rho)$.

Recalling that the assignation of probability is a function $p : \mathcal{D}(\mathcal{C}^2) \to [0,1]$, it is possible to establish the following equivalence relation in $\mathcal{D}(\mathbf{C}^2)$ given by

$$\sigma \equiv \nu \quad \text{if and only if} \quad p(\sigma) = p(\nu)$$

If we denote by $[\sigma]$ the equivalence class of $\sigma \in \mathcal{D}(\mathbf{C}^2)$, in view of Proposition 2.14 and Lemma 3.7 , it is clear that

$$[\sigma] = [(\sigma \bullet P_1)] = [\rho_{P(\sigma)}]$$

Thus we can consider the identification

$$(\mathcal{D}(\mathbf{C}^2)/\equiv) = (\rho_\lambda)_{\lambda \in [0,1]}$$

And the following proposition is easily demonstrated:

**Proposition 5.1** $\langle(\mathcal{D}(\mathbf{C}^2)/\equiv), \bullet, \oplus, \neg, [P_0], [P_1]\rangle$ *is a PMV-algebra isomorphic to* $[0,1]_{PMV}$. *The isomorphism is given by* $\rho_\lambda \to \lambda$. *The operations* $\vee, \wedge$ *give the PMV lattice structure in* $(\mathcal{D}(\mathbf{C}^2)/\equiv$. □

This proposition shows that a logic associated with quantum gates would admit an algebraic counterpart strongly associated with PMV-algebras. In other words, it would be a calculus based on the Łukaciewicz infinite valued calculus, enriched with a connective associated to **IAND**. This points a connection with fuzzy logic.

On the other hand, if we consider the finitely generated reduct of $\mathcal{D}(\mathbf{C}^2)$ given by $\langle P_0, P_1, \rho_{1/2} \bullet, \oplus, \neg\rangle$ we see that this reduct is a PMV-algebra isomorphic to the algebra $S$. In view of this, we can consider from a semantic level the Poincaré structure as equivalent to

$$\langle \mathcal{D}(\mathbf{C}^2), \bullet, \oplus, \neg, \sqrt{}, S\rangle$$

taking the elements of $S$ as constant operations. We adopt this last structure as a natural model of the propositional calculus associated with quantum gates. This propositional calculus will relate the properties of the probability assignation with respect to the operations associated to the gates that model the connectives. We will show that both the notion of theorem and of syntactic consequence in this logic system are equivalent to theorems of the theory of PMV propositional calculus, thus establishing the rigorous connection with fuzzy logic.

We will also consider the following problem: given a theory $T$ and a formula $\alpha$, we want to give a kind of parameter that shows the *relevance* *degree* of $T$ with respect to $\alpha$. It may be interpreted as a dependence factor of a possible output with respect to an input that is known with certainty. This reflects precisely what a quantum computer does.

For this development we adopt a Pavelka style calculus [13] in which the consideration of the elements of the subalgebra $S$ as constants will show its importance since they will be useful to "measure" the mentioned dependence factor. In other sense, this shows a strong relation with approximate reasoning, which is another item considered within fuzzy logic. Under this perspective, we will establish a completeness theorem.

We will call *Irreversible Quantum Computational Logic* the logic associated with the irreversible Poincaré structure, for short $IQCL$.

# 6 THE $IQCL$-PROPOSITIONAL CALCULUS

We introduce here the basic tools and properties associated with $IQCL$.

**Definition 6.1** The propositional $IQCL$-language, i.e. the notion of *formulas*, is defined in the usual way from the following alphabet

$$\langle P, \oplus, \odot, \rightarrow, \bullet, \vee, \wedge \neg, \surd, \overline{S}, (,) \rangle$$

where $P = \{p_1, p_2, ..., \}$ is the set of propositional variables, $\oplus, \odot, \rightarrow, \bullet, \vee, \wedge$ are binary connectives; $\neg, \surd$ are unary connectives, the set $\overline{S} = (\overline{s})_{s \in S}$ is the set of constant connectives representing the elements of the algebra $S$ and $(,)$ conform the punctuation symbols. We will refer to the elements of $P \cup \overline{S}$ as *atomic formulas*. In particular, we denote by $\bot, 1/2, \top \in \overline{S}$ the syntactic representatives of $P_o, \rho_{1/2}, P_1$ of the algebra $S$. In addition we introduce by definition the binary connective $\equiv$ as follows:

$$\alpha \equiv \beta \ \text{ for } \ (\alpha \rightarrow \beta) \odot (\beta \rightarrow \alpha)$$

We can establish the notion of complexity of formulas as a function $Comp : IQCL \rightarrow N$ such that $Comp(\alpha) = 0$ if $\alpha$ is an atomic formula; $Comp(\alpha * \beta) = 1 + Comp(\alpha) + Comp(\beta)$ for each binary conective $*$; $Comp(\neg\alpha) = Comp\sqrt{\alpha} = 1 + Comp(\alpha)$.

To begin considering the semantic aspect of this logic, we introduce the definition of *model*:

**Definition 6.2** A *model* of $IQCL$ is a function $e : \mathcal{L} \to \mathcal{D}(\mathcal{C}^2)$ such that satisfies the following conditions:

1. If $*$ is a binary connective, then $e(\alpha * \beta) = e(\alpha) * e(\beta)$
2. $e(\sqrt{\alpha}) = \sqrt{e(\alpha)}$
3. $e(\neg\alpha) = \neg e(\alpha)$
4. $e(\overline{s}) = \rho_s$ for each $\overline{s} \in (\overline{S})$
5. $e(\bot) = P_o$
6. $e(\top) = P_1$

It is clear that if two models $e$, $e'$ coincide over atomic formulas, then $e = e'$.

**Definition 6.3** 1. If $e : IQCL \to \mathcal{D}(\mathbf{C}^2)$ is a model and $\alpha \in IQCL$ then, the *probability-value* of $\alpha$ in $e$ is given by $e_p(\alpha) = p(e(\alpha))$.

2. $\beta$ *is a consequence of* $\alpha$ ($\alpha \models \beta$) if and only if $e_p(\alpha) \leq e_p(\beta)$ for each model $e$.

**Remark 6.4** It is not very hard to see that $e_p(\neg\alpha) = \neg e_p(\alpha)$ and if $*$ is a binary connective then $e_p(\alpha*\beta) = e_P(\alpha)*e_p(\beta)$. Because of the definition of *model* we can see that if $\overline{s} \in \overline{S}$ and $e^i, e^j$ are two models then $e^i_p(\overline{s}) = e^j_p(\overline{s})$. In view of this, for each $\overline{s} \in \overline{S}$ we refer to $p(\overline{s})$ as the probability-value of $\overline{s}$ since the probability value over $\overline{s}$ is independent of the model.

**Definition 6.5** A formula $\alpha$ is a *tautology* if and only if for each model $e$, $e_p(\alpha) = 1$

**Proposition 6.6** *Let $\alpha, \beta$ be formulas. Then we have*

$$\alpha \models \beta \quad \text{if and only if} \quad \alpha \to \beta \ \textit{is a tautology}$$

□

Thus, the idea of logical consequence is syntactically strongly related to the implicative connective.

**Definition 6.7** If we consider $\mathcal{D} = \{\rho = (0, r_2, r_3) : \rho \in \mathcal{D}(\mathbf{C}^2)\}$, then $e : IQCL \to \mathcal{D}$ are called *reduced models.*

**Proposition 6.8** *Let e be a model. Then there exists a reduced model $e'$ such that $e_p = e'_p$.*

*Proof:* By Proposition 3.2 the probability-values of formulas in $e$ depend on the coordinates $r_2, r_3$. If $\alpha$ is an atomic formula with $e(\alpha) = (r_1, r_2, r_3)$ we define $e'$ as $e'(\alpha) = (0, r_2, r_3)$. Finally we extend the model $e'$ by an inductive argument on the complexity of formulas, obtaining $e_p = e'_p$ □

**Remark 6.9** From the last proposition and since our semantics depends only on the probability-values of formulas, we can restrict the use of models to reduced models.

**Proposition 6.10** *Let $e, e'$ be two reduced models such that for each atomic formula $\alpha$, $e_p(\alpha) = e'_p(\alpha)$ and $e_p(\sqrt{\alpha}) = e'_p(\sqrt{\alpha})$. Then, we have $e = e'$.*

*Proof:* Let $\alpha$ be an atomic formula. Suppose that $e(\alpha) = (0, r_2, r_3)$, and $e'(\alpha) = (0, r'_2, r'_3)$. Since $e_p(\alpha) = e'_p(\alpha)$ and $e_p(\sqrt{\alpha}) = e'_p(\sqrt{\alpha})$ then, by Proposition 3.2, we have that $r_3 = r'_3$ and $r_2 = r'_2$. Finally, by an inductive argument on the complexity of formulas, it results that $e = e'$ □

**Definition 6.11** The following formulas are axioms of the $IQCL$:

*Lukasiewicz axioms:*

W1 $\alpha \to (\beta \to \alpha)$

W2 $(\alpha \to \beta) \to ((\beta \to \gamma) \to (\alpha \to \gamma))$

W3 $(\neg\alpha \to \neg\beta) \to (\beta \to \alpha)$

W4 $((\alpha \to \beta) \to \beta) \to ((\beta \to \alpha) \to \alpha)$

*Equivalence operations axioms:*

E1 $\alpha \odot \beta \equiv \neg(\neg\alpha \oplus \neg\beta)$

E2 $\alpha \to \beta \equiv \neg(\alpha \odot \neg\beta)$

E3 $\neg\alpha \equiv \alpha \to \bot$

E4 $\alpha \wedge \beta \equiv \alpha \odot (\alpha \to \beta)$

E5 $\alpha \vee \beta \equiv (\alpha \to \psi) \to \beta$

E6 $\neg\bot \equiv \top$

*Product axioms:*

P1 $(\alpha \bullet \beta) \to (\beta \bullet \alpha)$

P2 $(\top \bullet \alpha) \equiv \alpha$

P3 $(\alpha \bullet \beta) \to \beta$

P4 $(\alpha \bullet \beta) \bullet \gamma \equiv \alpha \bullet (\beta \bullet \gamma)$

P5 $x \bullet (y \odot \neg z) \equiv (x \bullet y) \odot \neg(x \bullet z)$

$S$ *axioms:* for each $\overline{r}, \overline{t} \in \overline{S}$

S1 $\overline{r} \odot \overline{t} \equiv \overline{r \odot t}$

S2 $\overline{r} \to \overline{t} \equiv \overline{r \to t}$

S3 $\overline{r} \bullet \overline{t} \equiv \overline{r \bullet t}$

$\sqrt{NOT}$ *axioms:*

Q1 $\sqrt{\sqrt{\alpha}} \equiv \neg\alpha$

Q2 $\sqrt{\neg\alpha} \equiv \neg\sqrt{\alpha}$

Q3 If $*$ is a binary connective $\sqrt{\alpha * \beta} \equiv \overline{1/2}$

Q4 $\sqrt{\overline{s}} \equiv 1/2$ for each $\overline{s} \in \overline{S}$.

Q5 $\{((\overline{\tfrac{1}{4}} \bullet \alpha) \oplus ((\overline{\tfrac{1}{4}} \bullet \sqrt{\alpha})) \to \overline{s} : \ \ s \geq (1 + \sqrt{2})/4\sqrt{2}\}$

The unique deduction rule of $IQCL$ is the *modus ponens* (MP).

We note that the axiom $Q5$ refers to the relation between density operators $\sigma, \sqrt{\sigma}$ with respect to the probability values they may take, i.e. between components $r_2, r_3$ (See Lemma 3.7). It is not very hard to see that

**Lemma 6.12** *The axioms of IQCL are tautologies.* □

A *theory* over $IQCL$ is a set of formulas $T$. A *proof* in $T$ is a sequence of formulas $\alpha_1 ... \alpha_n$ such that each member is either an axiom of $IQCL$ or a member of $T$ or follows from some preceding members of the sequence using *modus ponens.* $T \vdash \alpha$ means that $\alpha$ is *provable* in $T$, that is, $\alpha$ is the last formula of a proof from $T$. If $T = \emptyset$, we use the symbology $\vdash \alpha$ and, in this case, we will say that $\alpha$ is a *theorem of IQCL.* A model $e$ of $IQCL$ is a *model of a theory* $T$ if and only if $e_p(\alpha) = 1$ for each $\alpha \in T$. In this case we will use the notation $e_p(T) = 1$. We use $T \models \alpha$ or *semantic consequence* in the case in which when $e_p(T) = 1$ then $e_p(\alpha) = 1$ for each model $e$.

The following proposition is easily verified from the fact that the inference rule MP preserves the probability value 1.

**Proposition 6.13** *Let $T$ be a theory in $IQCL$ and $e$ be a model. If $e_p(T) = 1$ and $T \vdash \varphi$ then $e_p(\varphi) = 1$* □

**Remark 6.14** Note that axioms W1,...,W4, E1,...,E6 conform the same propositional deductive system as the infinite valued Łukasiewicz calculus [13, Chapt 3.1]). We had to add the Equivalence axioms in order that our notion of model resulted well defined. In fact, if we had considered for example $\rightarrow$ and $\perp$ as primitive connectives and defined $\neg\alpha$ as $\alpha \rightarrow \perp$, then we would have had to require that $e_p(\sqrt{\neg\alpha}) = e_p(\sqrt{\alpha \rightarrow \perp}) = 1/2$. But this is not in general true except when $e(\alpha) = \rho_\lambda$ for some $\lambda$.

**Proposition 6.15** *Let $\alpha, \beta, \gamma \in IQCL$. Then we have*

*L1* $(\alpha \rightarrow \beta) \rightarrow ((\alpha \rightarrow \gamma) \rightarrow (\beta \rightarrow \gamma))$

*L2* $(\alpha \odot \beta) \rightarrow \alpha$

*L3* $(\alpha \odot \beta) \rightarrow (\beta \odot \alpha)$

*L4* $(\alpha \odot (\alpha \rightarrow \beta)) \rightarrow (\beta \odot (\beta \rightarrow \alpha))$

*L5* $(\alpha \rightarrow (\beta \rightarrow \gamma)) \rightarrow ((\alpha \odot \beta) \rightarrow \gamma)$

*L6* $((\alpha \odot \beta) \rightarrow \gamma) \rightarrow (\alpha \rightarrow (\beta \rightarrow \gamma))$

*L7* $((\alpha \rightarrow \beta) \rightarrow \gamma)) \rightarrow (((\beta \rightarrow \alpha) \rightarrow \gamma) \rightarrow \gamma)$

*L8* $\bot \to \alpha$

*Proof:* Follows from W1...W4, E1...E6. and [13, Lemma 3.1.9, Corollary 3.1.16, and Definition 2.2.4 ] □

In view of Ł3 in the last proposition, we can introduce the following notation: for each formula $\alpha$, $\alpha^n$ is $\alpha \odot \cdots \odot \alpha$ (n-times).

**Proposition 6.16** *Let $\alpha, \beta \in IQCL$ and $T$ be a theory then we have*

1. $\vdash \alpha \to (\beta \to \alpha)$,
2. $\vdash \alpha \to (\beta \to (\alpha \odot \beta))$,
3. $\vdash (\alpha \to \beta) \to ((\alpha \odot \gamma) \to (\beta \odot \gamma))$,
4. $T \vdash \alpha \odot \beta$ *if and only if* $T \vdash \alpha$ *and* $T \vdash \beta$,
5. $T \vdash \alpha \equiv \beta$ *if and only if* $T \vdash \alpha \to \beta$ *and* $T \vdash \beta \to \alpha$,
6. $T \vdash \alpha \to \beta$ *and* $T \vdash \beta \to \gamma$ *then* $T \vdash \alpha \to \gamma$,
7. $\vdash \top$
8. $\vdash \alpha \to (\top \odot \alpha)$
9. $\vdash (\top \to \alpha) \to \alpha$
10. $\vdash \neg\neg\alpha \to \alpha$
11. $\vdash (\alpha \to \beta) \to (\neg\beta \to \neg\alpha)$,
12. $\vdash (\alpha \to \beta) \to ((\alpha \oplus \gamma) \to (\beta \oplus \gamma))$,
13. $\vdash ((\alpha \equiv \beta) \odot (\beta \equiv \gamma)) \to (\alpha \equiv \gamma)$
14. $\vdash (\alpha \equiv \beta) \to ((\alpha \to \gamma) \equiv (\beta \to \gamma))$
15. $\vdash (\alpha \equiv \beta) \to ((\gamma \to \alpha) \equiv (\gamma \to \beta))$
16. $\vdash (\alpha \to \beta)^n \vee (\beta \to \alpha)^n$ *for each* $n \in N$
17. $\vdash (\alpha \to \beta) \to ((\gamma \bullet \alpha) \to (\gamma \bullet \beta))$
18. $\vdash (\alpha \to \beta) \to ((\alpha \bullet \gamma) \to (\beta \bullet \gamma))$

*19.* $\vdash (\alpha \odot \beta) \to (\alpha \bullet \beta)$.

*Proof:* *1* see [13, Lemma 2.2.7]) *2, 3* see [13, Lemma 2.2.8]). *4* follows from item *2* and Ł2. *5* follows from item *4*. *6* follows from Ł1. *7, 8, 9* see [13, Lemma 2.2.14]). *10* see [13, Lemma 3.1.6]). *11* see [13, Lemma 2.2.12]). *12* straightforward from *3*, *11* and Axiom E1. *13, 14, 15* see [13, Lemma 2.2.16]). *16* see [13, Lemma 2.2.24]). Now we prove *17*

(1) $\vdash (\gamma \bullet (\alpha \odot \neg\beta)) \to ((\alpha \odot \neg\beta))$ by Ax P3

(2) $\vdash ((\gamma \bullet \alpha) \odot \neg(\gamma \bullet \beta)) \to (\gamma \bullet (\alpha \odot \neg\beta))$ by Ax P5, Prop 6.16 - 5

(3) $\vdash ((\gamma \bullet \alpha) \odot \neg(\gamma \bullet \beta)) \to (\alpha \odot \neg\beta)$ by 1,2 Prop 6.16 - 6

(4) $\vdash (((\gamma\bullet\alpha)\odot\neg(\gamma\bullet\beta)) \to (\alpha\odot\neg\beta)) \to (\neg(\alpha\odot\neg\beta) \to \neg((\gamma\bullet\alpha)\odot\neg(\gamma\bullet\beta)))$ by Prop 6.16 - 11

(5) $\vdash \neg(\alpha \odot \neg\beta) \to \neg((\gamma \bullet \alpha) \odot \neg(\gamma \bullet \beta))$ by MP 3,4

(6) $\vdash (\alpha \to \beta) \to \neg(\alpha \odot \neg\beta)$ by Ax E2, Prop 6.16 - 5

(7) $\vdash (\alpha \to \beta) \to \neg((\gamma \bullet \alpha) \odot \neg(\gamma \bullet \beta))$ by 5,6 Prop 6.16 - 6

(8) $\vdash \neg((\gamma \bullet \alpha) \odot \neg(\gamma \bullet \beta)) \to ((\gamma \bullet \alpha) \to (\gamma \bullet \beta))$ by Ax E2, Prop 6.16 - 5

(9) $\vdash (\alpha \to \beta) \to ((\gamma \bullet \alpha) \to (\gamma \bullet \beta))$ by 7,8 and Prop. 6.16-6

*18* Follows from 13, 14, 15, 17. Now we prove *19.*

(1) $\vdash (\top \to \alpha) \to ((\top \bullet \beta) \to (\alpha \bullet \beta))$ by Prop 6.16 - 17

(2) $\vdash ((\top \to \alpha) \to ((\top \bullet \beta) \to (\alpha \bullet \beta))) \to (((\top \to \alpha) \odot (\top \bullet \beta)) \to (\alpha \bullet \beta))$ by Ł5

(3) $\vdash ((\top \to \alpha) \odot (\top \bullet \beta)) \to (\alpha \bullet \beta)$ by MP1, 2

(4) $\vdash \alpha \to (\top \to \alpha)$ by Prop 6.16 - 1

(5) $\vdash (\alpha \to (\top \to \alpha)) \to ((\alpha \odot (\top \bullet \beta)) \to ((\top \to \alpha) \odot (\alpha \bullet \beta)))$ by Prop 6.16 - 3

(6) $\vdash (\alpha \odot (\top \bullet \beta)) \to ((\top \to \alpha) \odot (\top \bullet \beta)))$ by MP 4, 5

(7) $\vdash (\alpha \odot (\top \bullet \beta)) \to (\alpha \bullet \beta)$ by 3,6, Prop 6.16 - 6

(8) $\vdash \beta \to (\top \bullet \beta)$ by Ax P2, Prop 6.16 - 5

(9) $\vdash (\beta \to (\top \bullet \beta)) \to ((\alpha \odot \beta) \to (\alpha \odot (\top \bullet \beta))$ by Prop 6.16 - 3

(10) $\vdash (\alpha \odot \beta) \to (\alpha \odot (\top \bullet \beta))$ by MP 8,9

(11) $\vdash (\alpha \odot \beta) \to (\alpha \bullet \beta)$ by 7, 10, Prop 6.16 - 6

□

Now we can establish a kind of deduction theorem for the $IQCL$ calculus:

**Theorem 6.17** *Let $T$ be a theory and $\alpha$, $\beta$ be formulas. Then we have*

$$\text{if } T \cup \{\alpha\} \vdash \beta \text{ then there exists } n \in N \text{ such that } T \vdash \alpha^n \to \beta$$

*Proof:* See [13, Lemma 2.2.18]) □

**Definition 6.18** A theory $T$ is *inconsistent* if and only if $T \vdash \bot$; otherwise it is *consistent.*

**Lemma 6.19** *Let $T$ be theory. Then we have*

1. *$T$ is inconsistent if and only if $T \vdash \alpha$ for each formula $\alpha$,*
2. *$T \cup \{\alpha\}$ is inconsistent if and only if $T \vdash \neg\alpha^n$ for some $n \in N$*
3. *If $T$ is consistent then $T$ does no prove $\overline{s}$ for each $\overline{s} \in \overline{S} - \{\top\}$.*

*Proof:* 1) See [13, Lemma 2.2.21]). 2) See [13, Lemma 2.2.22]). 3)Follows from the axiom S1, Proposition 4.5 and Proposition 4.12. □

**Definition 6.20** Let $T$ be a theory over $IQCL$ and $\alpha$ be a formula

1 The *relevance degree* of T over $\alpha$ is $\|\alpha\|_T = \bigwedge\{e_p(\alpha) : e_p(T) = 1\}$

2 The *proof degree* of $\alpha$ is $|\alpha|_T = \bigvee\{r \in S : T \vdash \overline{r} \to \alpha\}$

**Remark 6.21** As we have already mentioned in the previous section, this relevance degree (also called *truth degree* in the frame of Pavelka style logics) represents the semantic relevance of a theory $T$ with respect to a formula $\alpha$, that is, the greatest lower bound of the probability values that $\alpha$ may take if all probability values of the formulas of $T$ are known to be 1. Our purpose is to show the equality between relevance degree and proof degree, the *Pavelka style strong completeness theorem.* This equality will give a syntactic notion of the relevance degree in the sense that it is the lowest upper bound $r \in S$ such that $T \vdash \overline{r} \to \alpha$.

**Lemma 6.22** *Let $T$ be a theory and $\alpha$ be a formula. Then we have:*

1. *If $T$ admits a model, then $\|\bot\|_T = 0$. In the other case, $\|\bot\|_T = 1$,*
2. *If $T$ is inconsistent, then $\|\alpha\|_T = 1$ for each formula $\alpha$,*
3. *If $s \in S$ is such that $s < |\alpha|_T$ then $T \vdash \overline{s} \to \alpha$.*

*Proof:* 1)If $T$ admits a model the result is obvious. In the other case, $\{e_p(\bot) : e_p(T) = 1\} = \emptyset$ resulting $\|\bot\|_T = 1$. 2) If $T$ is inconsistent, then $T \vdash \bot$, so $T$ does not admit a model. Thus for each formula $\alpha$, $\{e_p(\alpha) : e_p(T) = 1\} = \emptyset$, resulting $\|\alpha\|_T = 1$. 3) If $s < |\alpha|_T$ then, by definition of $|\alpha|_T$, there exists $s_1 \in S$ such that $s < s_1 \leq |\alpha|_T$ and $T \vdash \overline{s_1} \to \alpha$. By axioms S1...S3 it is not very hard to see that $\vdash \overline{s} \to \overline{s_1}$. Finally, by Proposition 6.15-6, $T \vdash \overline{s} \to \alpha$ □

# 7 THE PMV-FRAGMENT OF $IQCL$

In order to establish the formal connection between $IQCL$ and fuzzy logic via a Pavelka style strong completeness theorem, we will deal with a fragment of $IQCL$ whose algebraic counterpart is a PVM-algebra so to be able to use the results obtained in Section 4.

**Definition 7.1** We build the PMV-fragment, noted $IQCL_{PMV}$, from the sub language of $IQCL$ given by

$$\langle P \cup (\sqrt{p})_{p \in P}, \;\; \oplus, \odot, \rightarrow, \bullet, \wedge, \vee, \neg, \overline{S}, (,) \rangle$$

We consider the subsystem of $IQCL$ given by the axioms W1...W4, E1...E6, P1...P5, S1...S3 as the axiomatic system for $IQCL_{PMV}$.

In this section we must take the formulas $(\sqrt{p})_{p\in P}$ as atomic formulas of the fragment. We denote by $\vdash_{PMV}$ deductions on the PMV-fragment. A valuation over $IQCL_{PMV}$ is a function $v : IQCL_{PMV} \to [0,1]_{PMV}$ such that $v(\alpha * \beta) = v(\alpha) * v(\beta)$ for each binary connective $*$, $v(\neg\alpha) = \neg v(\alpha)$, and $v(\overline{s}) = s$ for each $\overline{s} \in \overline{S}$. Let $T$ be a theory and $\alpha$ be a formula, both in $IQCL_{PMV}$. A formula $\alpha$ is called *$IQCL_{PMV}$-tautology* if and only if for each valuation $v$, $v(\alpha) = 1$. We define *relevance degree* in $IQCL_{PMV}$, $\|\alpha\|_T^{PMV}$ as in $IQCL$ but in terms of the valuation over $IQCL_{PMV}$. In the same way the *proof degree* $|\alpha|_T^{PMV}$ is defined in the PMV-fragment taking the axiomatic system of $IQCL_{PMV}$.

**Remark 7.2** Note that Proposition 6.15 and 6.16 are also valid in the PMV-fragment.

**Definition 7.3** Let $T$ be a theory in $IQCL_{PMV}$. Then $T$ is said to be *complete* if and only if, for each pair of formulas $\alpha, \beta$ we have: $T \vdash_{PMV} \alpha \to \beta$ or $T \vdash_{PMV} \beta \to \alpha$.

**Lemma 7.4** *Let $T$ be a theory and $\alpha$ be a formula, both in $IQCL_{PMV}$. Suppose that $T$ does not prove $\alpha$ in the PMV-fragment. Then there exists a consistent complete theory $T'$ in $IQCL_{PMV}$ such that $T \subseteq T'$ and $T'$ does not prove $\alpha$ in the PMV-fragment.*

*Proof:* See [13, Lemma 2.4.2]) □

**Theorem 7.5** *Let $T$ be theory over $IQCL_{PMV}$. For each formula $\alpha$ we define $[\alpha] = \{\beta : T \vdash_{PVM} \alpha \equiv \beta\}$. Let $L_T = \{[\alpha] : \alpha \in IQCL_{PMV}\}$. If we define the following operations in $L_T$:*

$$0 = [\bot]$$

$$1 = [\top]$$

$$\neg[\alpha] = [\neg\alpha]$$

$$[\alpha] * [\beta] = [\alpha * \beta] \textit{ for each binary connective } *$$

*then we have*

1. *$\langle L_T, \ \oplus, \odot, \to, \bullet, \wedge, \vee, \neg, 0, 1\rangle$ is a PMV-algebra and $([\overline{s}])_{s\in S}$ is a sub algebra isomorphic to the algebra $S$.*

*2. If $T$ is a complete theory then $L_T$ is totally ordered.*

*Proof:* We first prove that the operations are well defined on $L_T$. In the case of $\oplus, \odot, \rightarrow, \wedge, \vee, \neg$ we refer to [13, Lemma 2.3.12]. Then we prove the well definition of $\bullet$. Suppose that $[\alpha_1] = [\alpha_2]$ and $[\beta_1] = [\beta_2]$, we need to see that $[\alpha_1] \bullet [\beta_1] = [\alpha_2] \bullet [\beta_2]$ or equivalently $[\alpha_1 \bullet \beta_1] = [\alpha_2 \bullet \beta_2]$

| | | |
|---|---|---|
| (1) | $T \vdash_{PMV} \alpha_1 \rightarrow \alpha_2$ | by Hyp, Prop. 6.16 5 |
| (2) | $T \vdash_{PMV} \beta_1 \rightarrow \beta_2$ | by Hyp, Prop. 6.16 5 |
| (3) | $\vdash_{PMV} (\alpha_1 \rightarrow \alpha_2) \rightarrow ((\alpha_1 \bullet \beta_1) \rightarrow (\alpha_2 \bullet \beta_2))$ | by Prop. 6.16 16 |
| (4) | $T \vdash_{PMV} (\alpha_1 \bullet \beta_1) \rightarrow (\alpha_2 \bullet \beta_2)$ | by MP 2, 3 |
| (5) | $\vdash_{PMV} (\beta_1 \rightarrow \beta_2) \rightarrow ((\alpha_2 \bullet \beta_1) \rightarrow (\alpha_2 \bullet \beta_2))$ | by Prop. 6.16 16 |
| (6) | $T \vdash_{PMV} ((\alpha_2 \bullet \beta_1) \rightarrow (\alpha_2 \bullet \beta_2)$ | by MP 4, 5 |
| (7) | $T \vdash_{PMV} ((\alpha_1 \bullet \beta_1) \rightarrow (\alpha_2 \bullet \beta_2)$ | by Prop. 6.16 6 |

In the same way we can prove that $T \vdash_{PMV} (\alpha_2 \bullet \beta_2) \rightarrow (\alpha_1 \bullet \beta_1)$ resulting $[\alpha_1 \bullet \beta_1] = [\alpha_2 \bullet \beta_2]$, and thus the well definition of $\bullet$. By [13, Lemma 2.3.12], the reduct $\langle L_T, \ \oplus, \odot, \rightarrow, \wedge, \vee, \neg, 0, 1\rangle$ is an MV-algebra. In view of this, if we consider the axioms P1...P5, this algebra is also a PMV-algebra. By axioms S1...S3 it results immediately that $([\overline{s}])_{s \in S}$ is a sub algebra isomorphic to the algebra $S$. For the other items, we refer to [13, Lemma 2.4.2] taking into account Lemma 4.13. □

We will refer to algebra $L_T$ as the *Lindenbaum algebra* associated to the theory $T$. Now we will establish a Pavelka style strong completeness theorem for the PMV-fragment:

**Theorem 7.6** *Let $T$ be a theory and $\alpha$ be a formula, both over $IQCL_{PMV}$. Then we have*

$$|\alpha|_T = \|\alpha\|_T$$

*Proof:* If $T$ is inconsistent, this result is trivial. Assume that $T$ is consistent. We first prove that $|\varphi|_T$ is a lower bound of $\{v(\varphi) : v(T) = 1\}$. Let $v$ be a valuation such that $v(T) = 1$. If $r \in S$ such that $T \vdash \overline{r} \rightarrow \varphi$, then we have that $1 = v(\overline{r} \rightarrow \varphi) = r \rightarrow v(\varphi)$. Thus $r \leq e(\varphi)$ resulting

$|\varphi|_T \leq v(\varphi)$. We proceed now to prove that $|\varphi|_T$ is the greatest lower bound of $\{v(\varphi) : v(T) = 1\}$. In fact, let $b$ be a lower bound of $\{v(\varphi) : v(T) = 1\}$. Suppose that $|\varphi|_T < b$. By Proposition 4.11, there exists $r_0 \in S$ such that $|\varphi|_T < r_0 < b$. Thus $T$ does not prove $\overline{r_0} \to \varphi$. Then, by Proposition 7.4, there exists a complete consistent theory $T'$ such that $T \subseteq T'$ and $T'$ does not prove $\overline{r_0} \to \varphi$. By Proposition 7.5, $L_{T'}$ is a totally ordered PMV-algebra containing $([\overline{s}])_{r \in S}$ as sub algebra isomorphic to $S$. Since $T'$ does not prove $\overline{r_0} \to \varphi$, then $[\overline{r}] \to [\varphi] < 1$ in $L_{T'}$, and then $[\varphi] < [\overline{r}_0]$ since it is totally ordered. Let $i_1$ be the canonical embedding $([\overline{s}])_{r \in S} \to S \subseteq [0,1]_{PMV}$ and $i_2$ be the canonical embedding $([\overline{s}])_{r \in S} \to L'_T$. Since $[0,1]_{PMV}$ is an injective object in the variety $\mathcal{PMV}$, there exists an homomorphism $f : L_{T'} \to [0,1]_{PVM}$ such that the following diagram is commutative

$$\begin{array}{ccc} \overline{S} & \xrightarrow{i_1} & [0,1]_{PVM} \\ {\scriptstyle i_2}\downarrow & \equiv & \nearrow {\scriptstyle f} \\ L_{T'} & & \end{array}$$

By commutativity of the diagram, $f([\varphi]) \leq r_0 < b$. If we consider the natural projection $\pi : IQCL_{PVM} \to L_{T'}$ such that $\alpha \to [\alpha]$, then the composition $f\pi$ is a valuation over $IQCL_{PVM}$ such that $f\pi(T') = 1$ and then $f\pi(T) = 1$ since $T \subseteq T'$ resulting $f\pi(\varphi) \in \{v(\varphi) : e(T) = 1\}$. But $f\pi(\varphi) \leq r_0 < b$ which is a contradiction since $b$ is a lower bound of $\{v(\varphi) : v(T) = 1\}$. Therefore $b \leq |\varphi|_T$ resulting $|\varphi|_T = \|\varphi\|_T$. □

# 8 COMPLETENESS OF $IQCL$

The study of the completeness of $IQCL$ is performed from the strong completeness of the PMV-fragment using a translation of formulas from $IQCL$ to $IQCL_{PMV}$.

**Definition 8.1** We define the PMV-*translation* as the function $\alpha \xrightarrow{t} \alpha_t$ such that:

$p \xrightarrow{t} p$ and $\sqrt{p} \xrightarrow{t} \sqrt{p}$ for each $p$ atomic formula.

$\sqrt{\neg\alpha} \xrightarrow{t} (\neg\sqrt{\alpha})_t$

$\sqrt{\sqrt{\alpha}} \xrightarrow{t} (\neg\alpha)_t$

$\sqrt{\alpha * \beta} \xrightarrow{t} \overline{1/2}$ for each binary connective $*$

$\alpha * \beta \xrightarrow{t} \alpha_t * \beta_t$ for each binary connective $*$

$\neg\alpha \xrightarrow{t} \neg(\alpha_t)$

If $T$ is a theory in $IQCL$, we define the PMV-translation over the theory as the set $T_t = \{\alpha_t : \alpha \in T\}$.

By a simple induction on complexity of formulas, it is not very hard to see that the PMV-translation is a function $t : IQCL \to IQCL_{PMV}$. From the definition of the PMV-translation, we can immediately establish the following lemma:

**Lemma 8.2** *Let $\alpha$ be a formula in $IQCL$. Then we have*

$$\vdash_{IQCL} \alpha \equiv \alpha_t$$

□

Taking into account the axiom $Q5$ of $IQCL$, we define the following theory in $IQCL_{PMV}$ which plays an important role in relation to deductions on $IQCL$ with respect to deductions in $IQCL_{PMV}$:

**Definition 8.3** We consider five groups of formulas in $IQCL_{PMV}$

$$T_1 = \{((\overline{\tfrac{1}{4}} \bullet p) \oplus ((\overline{\tfrac{1}{4}} \bullet \sqrt{p})) \to \overline{s} : p \in P,\ \ s \geq (1+\sqrt{2})/4\sqrt{2}\}$$

$$T_2 = \{((\overline{\tfrac{1}{4}} \bullet \neg p) \oplus ((\overline{\tfrac{1}{4}} \bullet \neg\sqrt{p})) \to \overline{s} : p \in P,\ \ s \geq (1+\sqrt{2})/4\sqrt{2}\}$$

$$T_3 = \{((\overline{\tfrac{1}{4}} \bullet \neg p) \oplus ((\overline{\tfrac{1}{4}} \bullet \sqrt{p})) \to \overline{s} : p \in P,\ \ s \geq (1+\sqrt{2})/4\sqrt{2}\}$$

$$T_4 = \{((\overline{\tfrac{1}{4}} \bullet p) \oplus ((\overline{\tfrac{1}{4}} \bullet \neg\sqrt{p})) \to \overline{s} : p \in P,\ \ s \geq (1+\sqrt{2})/4\sqrt{2}\}$$

$$T_5 = \{((\overline{\tfrac{1}{4}} \bullet \alpha) \oplus \overline{\tfrac{1}{8}}) \to \overline{s} : s \geq 3/8 = 1/4 \oplus 1/8\}$$

Then we define $T_{Q5} = T_1 \cup T_2 \cup T_3 \cup T_4 \cup T_5$.

**Lemma 8.4** *Let $\alpha \in IQCL$ and $s \in S$.*

*If $s \geq (1+\sqrt{2})/4\sqrt{2}\}$ then*

*1* $T_{Q5} \vdash_{PMV} (((\overline{\tfrac{1}{4}} \bullet \alpha) \oplus (\overline{\tfrac{1}{4}} \bullet \sqrt{\alpha})) \to \overline{s})_{\,t}$ *noted* $T_{Q5} \vdash_{PMV} \alpha^1_t$

*2* $T_{Q5} \vdash_{PMV} (((\overline{\tfrac{1}{4}} \bullet \neg\alpha) \oplus (\overline{\tfrac{1}{4}} \bullet \neg\sqrt{\alpha})) \to \overline{s})_{\,t}$ *noted* $T_{Q5} \vdash_{PMV} \alpha^2_t$

*3* $T_{Q5} \vdash_{PMV} (((\overline{\tfrac{1}{4}} \bullet \neg\alpha) \oplus (\overline{\tfrac{1}{4}} \bullet \sqrt{\alpha})) \to \overline{s})_{\,t}$ *noted* $T_{Q5} \vdash_{PMV} \alpha^3_t$

*4* $T_{Q5} \vdash_{PMV} (((\overline{\tfrac{1}{4}} \bullet \alpha) \oplus (\overline{\tfrac{1}{4}} \bullet \neg\sqrt{\alpha})) \to \overline{s})_{\,t}$ *noted* $T_{Q5} \vdash_{PMV} \alpha^4_t$

*If* $s \geq 3/8 = \overline{1/4} \oplus 1/8$

*5* $T_{Q5} \vdash_{PMV} (((\overline{\tfrac{1}{4}} \bullet \alpha) \oplus \overline{\tfrac{1}{8}}) \to \overline{s})_{\,t}$ *noted* $T_{Q5} \vdash_{PMV} \alpha^5_t$

*Proof:* We use induction on complexity of $\alpha$. If $Comp(\alpha) = 0$, we consider two cases: *Case* $\alpha \in \overline{S}$: By axiom Q4, $\sqrt{\alpha} \equiv \overline{1/2}$ and by axiom S2, $\neg\overline{1/2}. \equiv \overline{1/2}$. From this and using Proposition 6.16 the result is immediate. *Case* $\alpha \in P$: the translation of five formulas is over $T_{Q5}$. Suppose that this result is valid when $Comp(\alpha) < n$. Let $\alpha$ be such that $Comp(\alpha) = n$

Suppose $\alpha$ is $\neg\beta$.

1. $\alpha^1_t = (((\overline{\tfrac{1}{4}} \bullet \neg\beta) \oplus (\overline{\tfrac{1}{4}} \bullet \sqrt{\neg\beta})) \to \overline{s})_{\,t} = ((\overline{\tfrac{1}{4}} \bullet \neg\beta_t) \oplus (\overline{\tfrac{1}{4}} \bullet \neg(\sqrt{\beta})_t)) \to \overline{s} =$
   $(((\overline{\tfrac{1}{4}} \bullet \neg\beta) \oplus (\overline{\tfrac{1}{4}} \bullet \neg(\sqrt{\beta}))) \to \overline{s})_{\,t} = \beta^2_t$.
   By inductive hypothesis $T_{Q5} \vdash_{PMV} \beta^2_t$.

2. $\alpha^2_t = (((\overline{\tfrac{1}{4}} \bullet \neg\neg\beta) \oplus (\overline{\tfrac{1}{4}} \bullet \neg\sqrt{\neg\beta})) \to \overline{s})_{\,t} =$
   $((\overline{\tfrac{1}{4}} \bullet \neg\neg\beta_t) \oplus (\overline{\tfrac{1}{4}} \bullet \neg\neg(\sqrt{\beta})_t)) \to \overline{s}$
   But using Proposition 6.16
   $\vdash_{PMV} ((\overline{\tfrac{1}{4}} \bullet \neg\neg\beta_t) \oplus (\overline{\tfrac{1}{4}} \bullet \neg\neg(\sqrt{\beta})_t)) \to \overline{s} \equiv ((\overline{\tfrac{1}{4}} \bullet \beta_t) \oplus (\overline{\tfrac{1}{4}} \bullet (\sqrt{\beta})_t)) \to \overline{s}$
   being $((\overline{\tfrac{1}{4}} \bullet \beta_t) \oplus (\overline{\tfrac{1}{4}} \bullet (\sqrt{\beta})_t)) \to \overline{s} = \beta^1_t$. By inductive hypothesis we have $T_{Q5} \vdash_{PMV} \beta^1_t$.

3. $\alpha^3_t = (((\overline{\tfrac{1}{4}} \bullet \neg\neg\beta) \oplus (\overline{\tfrac{1}{4}} \bullet \sqrt{\neg\beta})) \to \overline{s})_{\,t} = ((\overline{\tfrac{1}{4}} \bullet \neg\neg\beta_t) \oplus (\overline{\tfrac{1}{4}} \bullet \neg(\sqrt{\beta})_t)) \to \overline{s}$
   But using Proposition 6.16
   $\vdash_{PMV} ((\overline{\tfrac{1}{4}} \bullet \neg\neg\beta_t) \oplus (\overline{\tfrac{1}{4}} \bullet \neg(\sqrt{\beta})_t)) \to \overline{s} \equiv ((\overline{\tfrac{1}{4}} \bullet \beta_t) \oplus (\overline{\tfrac{1}{4}} \bullet \neg(\sqrt{\beta})_t)) \to \overline{s}$
   being $((\overline{\tfrac{1}{4}} \bullet \beta_t) \oplus (\overline{\tfrac{1}{4}} \bullet \neg(\sqrt{\beta})_t)) \to \overline{s} = \beta^4_t$. By inductive hypothesis we have $T_{Q5} \vdash_{PMV} \beta^4_t$.

4. $\alpha_t^4 = (((\overline{\frac{1}{4}} \bullet \neg\beta) \oplus (\overline{\frac{1}{4}} \bullet \neg\sqrt{\neg\beta})) \to \overline{s})_t = ((\overline{\frac{1}{4}} \bullet \neg\beta_t) \oplus (\overline{\frac{1}{4}} \bullet \neg\neg(\sqrt{\beta})_t)) \to \overline{s}$
   But using Proposition 6.16
   $\vdash_{PMV} ((\overline{\frac{1}{4}} \bullet \neg\beta_t) \oplus (\overline{\frac{1}{4}} \bullet \neg\neg(\sqrt{\beta})_t)) \to \overline{s} \equiv ((\overline{\frac{1}{4}} \bullet \neg\beta_t) \oplus (\overline{\frac{1}{4}} \bullet (\sqrt{\beta})_t)) \to \overline{s}$
   being $((\overline{\frac{1}{4}} \bullet \neg\beta_t) \oplus (\overline{\frac{1}{4}} \bullet (\sqrt{\beta})_t)) \to \overline{s} = \beta_t^3$. By inductive hypothesis we have $T_{Q5} \vdash_{PMV} \beta_t^3$.

5. Immediate, since it results an element of $T_5$

Suppose $\alpha$ is $\sqrt{\beta}$.

1. $\alpha_t^1 = (((\overline{\frac{1}{4}} \bullet \sqrt{\beta}) \oplus (\overline{\frac{1}{4}} \bullet \sqrt{\sqrt{\beta}})) \to \overline{s})_t = ((\overline{\frac{1}{4}} \bullet \sqrt{\beta_t}) \oplus (\overline{\frac{1}{4}} \bullet \neg\beta_t)) \to \overline{s}$
   But using Proposition 6.16
   $\vdash_{PMV} ((\overline{\frac{1}{4}} \bullet \sqrt{\beta_t}) \oplus (\overline{\frac{1}{4}} \bullet \neg\beta_t)) \to \overline{s} \equiv ((\overline{\frac{1}{4}} \bullet \neg\beta_t) \oplus (\overline{\frac{1}{4}} \bullet \sqrt{\beta_t})) \to \overline{s}$
   being $((\overline{\frac{1}{4}} \bullet \neg\beta_t) \oplus (\overline{\frac{1}{4}} \bullet \sqrt{\beta_t})) \to \overline{s} = \beta_t^3$. By inductive hypothesis we have $T_{Q5} \vdash_{PMV} \beta_t^3$.

   For the rest of this case, it follows in a similar manner.

Suppose that $\alpha$ is $\beta * \gamma$ for $*$ a binary connective. Then we use the same procedure as for the atomic case for elements of $\overline{S}$.

□

The following theorem establishes the relation between the deductive system of $IQCL$ and the deductive system $IQCL_{PMV}$:

**Theorem 8.5** *Let $T$ be a theory and $\alpha$ be a formula both in $IQCL$. Then we have*

$$T \vdash_{IQCL} \alpha \quad \text{if and only if} \quad T_t \cup T_{Q5} \vdash_{PMV} \alpha_t$$

*Proof:*
Suppose that $T \vdash_{IQCL} \alpha$. We use induction on the length of the poof of $\alpha$ noted by $Length(\alpha)$. If $Length(\alpha) = 1$, then we have the following possibilities:

1. $\alpha$ is one of axioms W1, $\cdots$, W4, E1, $\cdots$, E6, P1, $\cdots$, P5, S1, $\cdots$, S3. In this case $\alpha_t$ result an axiom of the $IQCL_{PMV}$.

2. $\alpha$ is one of the axioms Q1, $\cdots$, Q4. In this case $\alpha_t$ looks like $\beta \equiv \beta$ in $IQCL_{PMV}$ and, by Proposition 6.16 1 and 9, these formulas are PMV-theorems.

3. If $\alpha$ is axiom Q5, then we use Lemma 8.4 resulting $T_{Q5} \vdash \alpha_t$

4. If $\alpha \in T$, it is clear that $\alpha_t \in T_t$.

Suppose that the theorem is valid for $Length(\alpha) < n$. We consider $Length(\alpha) = n$. Thus we have an $IQCL$-proof $\alpha$ from $T$ as follows

$$\alpha_1, \cdots, \alpha_m \to \alpha, \cdots, \alpha_m, \cdots, \alpha_{n-1}, \alpha$$

obtaining $\alpha$ by MP from $\alpha_m \to \alpha$ and $\alpha_m$. Using inductive hypothesis we have $T_t \cup T_{Q5} \vdash_{PMV} (\alpha_m \to \alpha)_t$ and $T_t \cup T_{Q5} \vdash_{PMV} (\alpha_m)_t$. Taking into account that $(\alpha_m \to \alpha)_t = (\alpha_m)_t \to \alpha_t$, by MP we have $T_t \cup T_{Q5} \vdash_{PMV} \alpha_t$. The converse is immediate from Lemma 8.2 and the fact that the formulas in $T_{Q5}$ are $IQCL$-theorems. $\square$

**Corollary 8.6** *Let $\alpha \in IQCL$, then we have*

$$\vdash_{IQCL} \alpha \quad \text{if and only if} \quad T_{Q5} \vdash_{PMV} \alpha_t$$

$\square$

**Remark 8.7** Theorem 8.5 shows in a formal sense the syntactic relation between $IQCL$ and fuzzy logic. More precisely, we recall that provable formulas from $IQCL$-theories are identifiable to provable formulas from PMV-theories obtained by translation plus $T_{Q5}$. Note that $IQCL$-theorems are PMV-theorems of $T_{Q5}$. Another important result that arises from these theorems is that the $\surd$ connective has "importance" only when applied to atomic formulas. More precisely, the real difficulty of $\surd$ in $IQCL$ is captured by the PMV-theory $T_{Q5}$.

Now we introduce the following sets in order to study the relation between reduced models of $IQCL$ and valuations of the PMV-fragment:

$E_R = \{e : e \text{ is a reduced model of } IQCL\}$

$V_{Q5} = \{v : v \text{ is a valuation of } IQCL_{PMV} \text{ with } v(T_{Q5}) = 1\}$

**Proposition 8.8** *There exists a bijection $E_R \to V_{Q5}$, $e \to v_e$, such that $e_p(\alpha) = v_e(\alpha_t)$.*

*Proof:* Let $e \in E_R$ and let $v_e = e_p | IQCL_{PMV}$. We will see that $e \to v_e$ is well defined in the sense that $v_e \in V_{Q5}$. Let $\alpha \in T_{Q5}$. If $\alpha \notin T_5$, then it is not very hard to see that the density operator $e(\alpha)$ has the form $e(\alpha) = (\frac{\sigma}{4} \oplus \frac{\sqrt{\sigma}}{4}) \to s$ for some density operator $\sigma$ and $s \geq \frac{1+\sqrt{2}}{4\sqrt{2}}$. But by Lemma 3.7, $p(\frac{\sigma}{4} \oplus \frac{\sqrt{\sigma}}{4}) \leq \frac{1+\sqrt{2}}{4\sqrt{2}}$. Thus $e_p(\alpha) = 1$. If $\alpha \in T_5$ we use the same argument with the bound $3/8$. Thus the assignation is well defined. The injectivity follows for Proposition 6.10. Now we will prove the surjectivity. Let $v \in V_{Q5}$. For each atomic formula $\alpha$ let $e(\alpha) = ((1 - 2v(\alpha)),\ (1 - 2v(\sqrt{\alpha})))$. Thus $e$ is well defined over atomic formulas since $v$ satisfies $T_{Q5}$. Then we can extend $e$ to $IQCL$ (with unicity, from Proposition 6.10) and it is clear that $e_p = v$ since $e_p$ and $v$ coincide over atomic formulas of $IQCL_{PMV}$. Finally, for each $IQCL$ formula $\alpha$, $e_p(\alpha) = v_e(\alpha_t)$ taking into account the inductive argument on the complexity of formulas and translation. □

**Remark 8.9** This last proposition shows the rigurous semantic connection between $IQCL$ and fuzzy logic. More precisely, models of $IQCL$ are identifiable to PMV-valuations that satisfy $T_{Q5}$.

**Proposition 8.10** *Let $T$ be a theory and $\alpha$ be a formula, both in $IQCL$. Then we have*

1. $|\alpha|_T^{IQCL} = |\alpha_t|_{T_t \cup T_{Q5}}^{PMV}$
2. $\|\alpha\|_T^{IQCL} = \|\alpha_t\|_{T_t \cup T_{Q5}}^{PMV}$

*Proof:* 1) Immediate from Theorem 8.5. 2) Follows from Proposition 8.8. □

Now we can establish a Pavelka style strong completeness theorem for $IQCL$.

**Theorem 8.11** *Let $T$ be a theory and $\alpha$ be a formula both in $IQCL$. Then we have*

$$|\alpha|_T^{IQCL} = \|\alpha\|_T^{IQCL}$$

*Proof:* Follows from Propsition 8.10 and Theorem 7.6.

□

**Corollary 8.12** *Let $T$ be a theory and $\alpha$ be a formula both in $IQCL$. If $\|\alpha\|_T^{IQCL} = 0$, then for each $s \in S$, $\overline{s} \to \alpha$ in not provable from $T$* □

From the above completeness theorem, we can establish a somehow compactness theorem.

**Theorem 8.13** *Let $T$ be a theory and $\alpha$ be a formula both in $IQCL$. Then we have:*

$$\text{If } r \leq \|\alpha\|_T^{IQCL} \text{ then } \exists\, T_0 \subseteq T \text{ finite such that } r \leq \|\alpha\|_{T_0}^{IQCL}$$

*Proof:* If $r \leq \|\alpha\|_T^{IQCL}$ then by Theorem 8.11 and Lemma 6.22 we have $T \vdash_{IQCL} \overline{r} \to \alpha$. If $\alpha_1, \cdots \alpha_n, \overline{r} \to \alpha$ is a proof of $\overline{r} \to \alpha$ from $T$, we can consider the finite set $T_0$ given as $T_0 = \{\alpha_k \in T : \alpha_k \in \{\alpha_1, \cdots \alpha_n\}\}$. Using again Theorem 8.11 we have $r \leq |\alpha|_{T_0}^{IQCL} = \|\alpha\|_{T_0}^{IQCL}$

□

# 9 CONCLUSIONS

The logical support associated with quantum treatment of information based on the Poincaré structure turns to be a particular case of the propositional theories related to fuzzy logic, more precisely to a type of infinite valued logic. This logical frame allows to use approximate reasoning of fuzzy logic as a tool for the study of the relevance that input data that are known with certainty have on the possible outputs.

### Aknowledgements

This work was partially supported by the following grants: PICT 04-17687 (ANPCyT), PIP N$^o$ 1478/01 (CONICET), UBACyT N$^o$ X081 and X204.

Graciela Domenech:
Instituto de Astronomía y Física del Espacio (IAFE)
Ciudad Universitaria
1428 Buenos Aires - Argentina
e-mail: domenech@iafe.uba.ar
Hector Freytes:
Escuela de Filosofía
Universidad Nacional de Rosario,
Entre Ríos 758, 2000, Rosario, Argentina
e-mail: hfreytes@dm.uba.ar